\documentclass[a4paper,11pt]{article}

\usepackage[titlenumbered,ruled]{algorithm2e}
\usepackage{fourier}
\usepackage{color}
\usepackage{graphicx}
\usepackage{url}
\usepackage[affil-it]{authblk}
\usepackage{amsmath}
\usepackage{wrapfig}
\usepackage{float}

\usepackage[T1]{fontenc}
\usepackage{times}

\usepackage{booktabs,caption}
\usepackage[flushleft]{threeparttable}
\usepackage{enumitem}
\usepackage{subfigure}
\DeclareUnicodeCharacter{2212}{-}
\newcommand{\comment}[1]{}
\newlength{\tempdima}

\newcommand{\rowname}[1]

\begin{document}

\title{
Random Data Augmentation based Enhancement: A Generalized Enhancement Approach for Medical Datasets }

\author[1]{Sidra Aleem}

\author[2]{Teerath Kumar}\author[2]{Suzanne Little} \author[2]{Malika Bendechache}

\author[3]{Rob Brennan}
\author[1]{Kevin McGuinness}

\affil[1]{School of Electronic Engineering, Dublin City University, Ireland}
\affil[2]{School of Computing, Dublin City University, Ireland}
\affil[3]{ADAPT, School of Computer Science, University College Dublin, Ireland}

\date{}
\maketitle
\thispagestyle{empty}

\begin{abstract}

Over the years, the paradigm of medical image analysis has shifted from manual expertise to automated systems, often using deep learning (DL) systems. The performance of deep learning algorithms is highly dependent on data quality. Particularly for the medical domain, it is an important aspect as medical data is very sensitive to quality and
poor quality can lead to misdiagnosis. To improve the diagnostic performance, research has been done both in complex DL architectures and in improving data quality using
dataset dependent static hyperparameters. However, the performance is still constrained due to data quality and overfitting of hyperparameters to a specific dataset. To overcome
these issues, this paper proposes random data augmentation based enhancement. The
main objective is to develop a generalized, data-independent and computationally efficient
enhancement approach to improve medical data quality for DL. The quality
is enhanced by improving the brightness and contrast of images. In contrast to the existing methods, our method generates enhancement
hyperparameters randomly within a defined range, which makes it robust and prevents overfitting to a specific dataset.
To evaluate the generalization of the proposed method, we
use four medical datasets and compare its performance with state-of-the-art methods for
both classification and segmentation tasks. For grayscale imagery, experiments have been
performed with: COVID-19 chest X-ray, KiTS19, and for RGB imagery with: LC25000 datasets. Experimental results demonstrate that with the proposed enhancement methodology, DL architectures outperform other existing methods.
Our code is publicly available at: \url{https://github.com/aleemsidra/Augmentation-Based-Generalized-Enhancement}.  
\end{abstract}
\textbf{Keywords:} Classification, Data Augmentation, Generalized Enhancement, Segmentation

\section{Introduction}\label{intro}
DL algorithms are being used in many domains  such image data~\cite{ kumar2021class}, audio data~\cite{chandio2021audd, turab2022investigating,kumar2020intra,park2020search} and many more~\cite{ kumar2021binary} and have revolutionized the field of medical image analysis. DL based applications are widely being used for computer assisted disease diagnosis to aid clinicians \cite{philipp2021localizing}. Regardless of the DL model architecture, performance is strongly affected by the quality of the raw data \cite{zhou2019contrast}. Particularly for medical images, data quality is a critical factor for reliable disease assessment and diagnosis \cite{chen2014review}. Further issues arise due to the differences of acquisition protocols and the heterogeneity of data \cite{boyat2015review}.
To work around these issues, research has been done both to introduce new DL architectures \cite{szegedy2015going, yadav2019deep, cai2020review} and to improve the data quality \cite{faes2019automated}. However, both of these solutions
pose further problems. The use of new complex DL architectures has a number of drawbacks: (a) the issue of data quality remains persistent, which affects the final prediction; (b) it does not give an insight to the actual performance capability of traditional DL architectures; (c) it increases computational complexity without addressing the actual quality issue. To improve data quality, the optimal selection of enhancement hyperparameters is extremely important. Due to the sensitive nature of medical data, this selection is even more important to have a robust performance independent of the data quality. The existing methods rely on fixed enhancement hyperparameters. These hyperparameters are chosen according to the dataset \cite{masud2021machine,mittal2019deep, hari2013unsharp}.
Consequently, such enhancement methods are prone to overfitting. Moreover, these techniques are evaluated either on grayscale or RGB datasets
\cite{gao2021reversible,wang2019variational, zhou2019contrast}. Thus, this the limited evaluation is not indicative of the strength of such methods. To prevent these issues, a computationally efficient and generalized contrast enhancement method can help.\par
To overcome above stated issues, this paper proposes random data augmentation based computationally efficient and generalized  enhancement method, in which data quality is improved using random brightness and contrast hyperparameters. 
Unlike other existing methods, which use fixed hyperparameters for enhancement, we use a set of hyperparameters. The hyperparameters are randomly chosen from this set and hence are not reliant on the dataset. The random selection ensures that the hyperparameters do not overfit the data. The data-independent nature of these hyperparameters aids the proposed method to generalize well on different datasets. The range of brightness set is $[1.15, 1.35]$ and the range of contrast set is $[−0.1, 0.4]$. 
These specific ranges are chosen by performing experimental evaluation. First we started from -1.0 for both hyperparameters, where images have apparently no features. We evaluated the corresponding affect on enhancement by visualizing resultant data and kept on  incrementing the value with interval of 0.15.  It was observed that images  started to show some feature at 1.15 and -0.1 brightness and contrast values respectively. Thus, these values were chosen as starting points for contrast and brightness sets. For the end point, we followed the same methodology and choose those values as end points before which images started to loose the features.   The enhancement results achieved with starting and end points of selected range are shown  in Figure \ref{fig:fig4}. Beyond this particular range, image becomes darker or brighter and starts losing features. The performance is assessed by evaluating the resultant enhanced data with a variety of traditional DL architectures.

\textbf{Contributions}: The main contributions of our work are as follows:
\begin{itemize}
  \setlength{\itemsep}{1pt}
  \setlength{\parskip}{0pt}
  \setlength{\parsep}{0pt}
    \item We propose a generalized and computationally efficient random data augmentation based enhancement approach for medical data.
    \item The enhancement hyperparameters are not manually selected according to the data; thus our enhancement method does not overfit a specific dataset.
    \item To check the effectiveness of our work, we perform  extensive experiments on both gray scale and RGB datasets for classification and segmentation tasks.
    \item The proposed approach shows superior performance in terms of both accuracy and execution time over state-of-the-art techniques.

\end{itemize}

\section{Related Work}
Generally three types of contrast enhancement methods have been used: histogram methods, spectral methods and spatial methods \cite{pierre2017variational}. The histogram methods have remained very popular for contrast enhancement. Such methods transform gray scale images to an image with a specified histogram. However, such methods result in poor enhancement that can be attributed to both loss of information and over-enhancement of specific gray levels. Such methods are not adaptive and thus are inappropriate to provide contrast enhancement for the medical imaging domain \cite{reddy2018retinal, singh2016contrast}. Spectral methods use wavelet transforms for quality enhancement. However, such methods fail to provide simultaneous enhancement to all the parts of of images. Moreover, it is difficult to automate enhancement using them \cite{wang2019variational}. The motive of the contrast enhancement in medical images is to aid
clinicians with automated diagnosis, so such methods are also not best suited for medical domain.\par
A combination of adaptive histogram equalisation and discrete wavelet transform for contrast enhancement was used in \cite{lidong2015combination}. This process is comprised of
three steps and gives a detailed output. However, the performance was affected due to contrast stretching and noise enhancement issues. Due to the sensitivity of medical domain to errors, disease diagnosis can be affected by such noise enhancement. Furthermore, the filter based methods need to find the appropriate filter and their performance is highly
reliant on the hyperparameters used. The hyperparameters are chosen according to the specific dataset. Thus these hyperparameters are not generalized to unseen datasets and consequently failed to perform well \cite{zhou2019contrast}. So, such methods will fail to cope
with the vast variety of medical data.\par
Gamma correction is one of the other enhancement methods used widely. Its performance is dependant on the $\gamma$ coefficient. To deal with the varying illumination, an adaptive gamma correction technique to modify two non-linear functions has been proposed in \cite{shi2007reducing}. However, these functions may be uniform
for various regions and patterns of an image. An adaptive gamma correction method based on cumulative distribution to modify the histogram has been proposed in \cite{huang2012efficient}.
\section{Proposed Method}
In this section, we introduce our proposed approach for random data augmentation based enhancement. Let $\alpha$ and $\beta$ be gain and bias. These parameters regulate contrast($\alpha$) and brightness ($\beta$). $f(i)$ is the input image and $g(i)$ is the resultant enhanced image \cite{steger2018machine} as shown in equation \ref {eq:eq1}:

\begin{equation}\label{eq:eq1}
g(i) = \alpha  f(i) + \beta.
\end{equation}
Unlike other enhancement methods, which use fixed hyperparameters for enhancement and are prone to overfitting on particular dataset, our method randomly picks the value for $\alpha$ and $\beta$ from the defined sets. To prevent overfitting, the set used consists of both positive and negative values for $\alpha$ and $\beta$. The values are randomly picked up from this set and are then used to augment data as shown in Algorithm 1. As $\alpha$ and $\beta$ are data independent, this makes our proposed method a generalized enhancement approach that is suitable for different types of datasets. 
\begin{figure}[!hb]
\begin{center} 
  {\includegraphics[width=0.4\linewidth, height =0.4\linewidth]{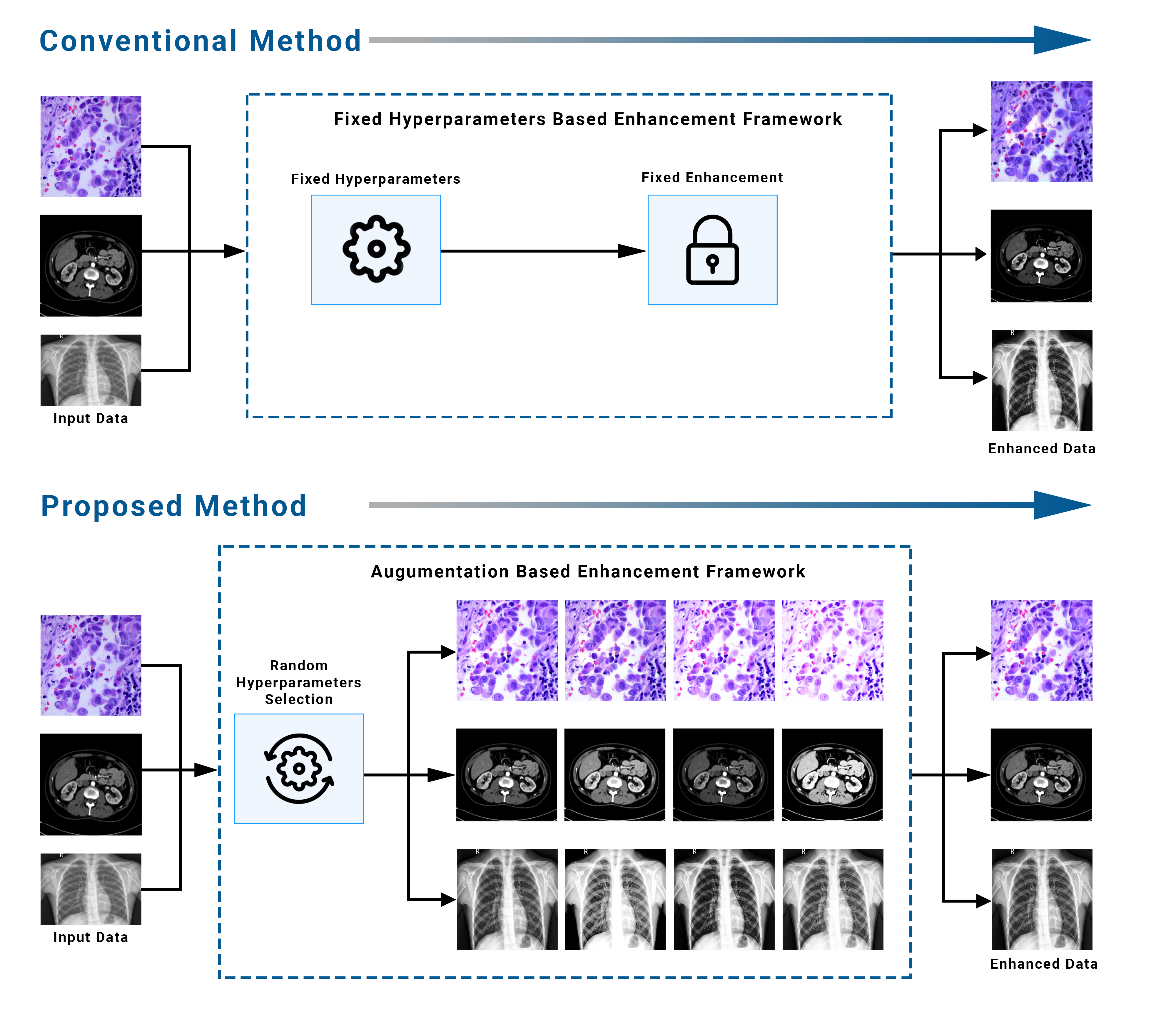}
  }
   {\caption{Overview of the proposed random augmentation based enhancement and the conventional enhancement}}
\end{center}
  \label{fig:work_flow}
\end{figure}

\begin{minipage}{1\linewidth}
\begin{algorithm}[H]
    \SetKwInOut{Input}{input}
    \SetKwInOut{Output}{output}
\Input{$f(i)$ : batch of images\newline
  $n$: batch size\newline
  alpha (gain): randomly generated  \newline 
  beta (bias): randomly generated}
    \Output{$g(i)$: Enhanced images batch}
    \caption{Random Data Augmentation based Enhancement}
    \label{alg:net}
    \For{$f(i)\leftarrow 1$  \KwTo $n$}{
      $random\_index \leftarrow randrange(len(alpha))$\\
      $r\_alpha \leftarrow alpha[random\_index]$\\
      $random\_index \leftarrow randrange(len(beta))$\\
      $r\_beta \leftarrow alpha[random\_index]$\\
      $g(i) = clip(r\_alpha * f(i) + r\_beta, 0,255)$\\
    }
\end{algorithm}
\end{minipage}

\section{Experimental Results}
\subsection{Datasets}
\begin{figure}[!hb]
    \centering
    \begin{subfigure}[]
    {\includegraphics[width=0.15\textwidth, height =0.12\textwidth]{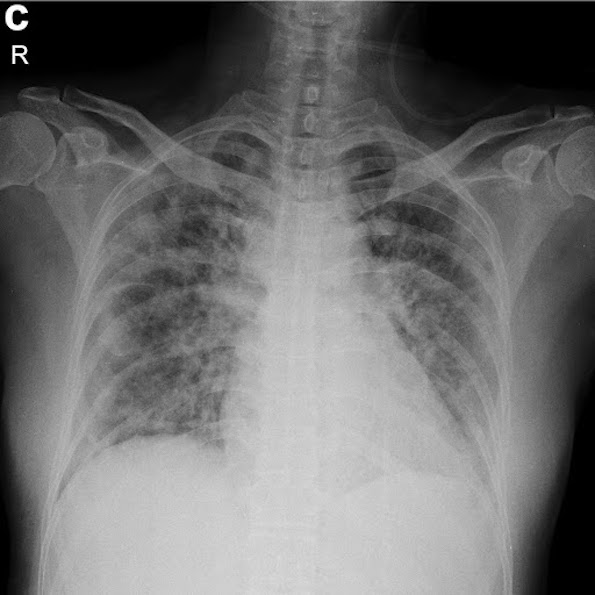}} 
    \end{subfigure}
    \begin{subfigure}[]
    {\includegraphics[width=0.15\textwidth, height =0.12\textwidth]{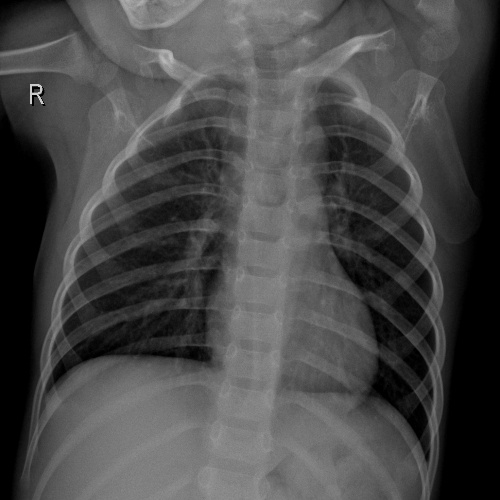}} 
    \end{subfigure}
     \begin{subfigure}[]
    {\includegraphics[width=0.15\textwidth, height =0.12\textwidth]{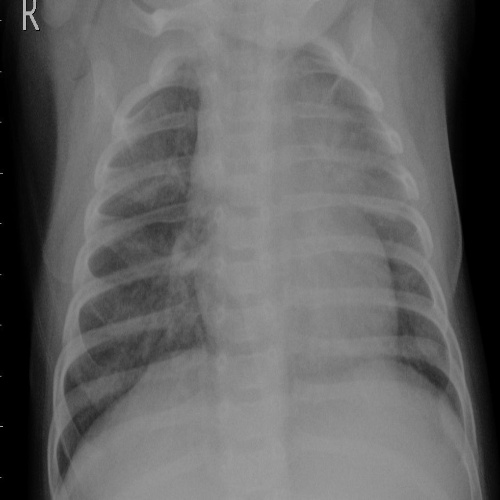}}
     \end{subfigure}
      \begin{subfigure}[]
    {\includegraphics[width=0.15\textwidth, height =0.12\textwidth]{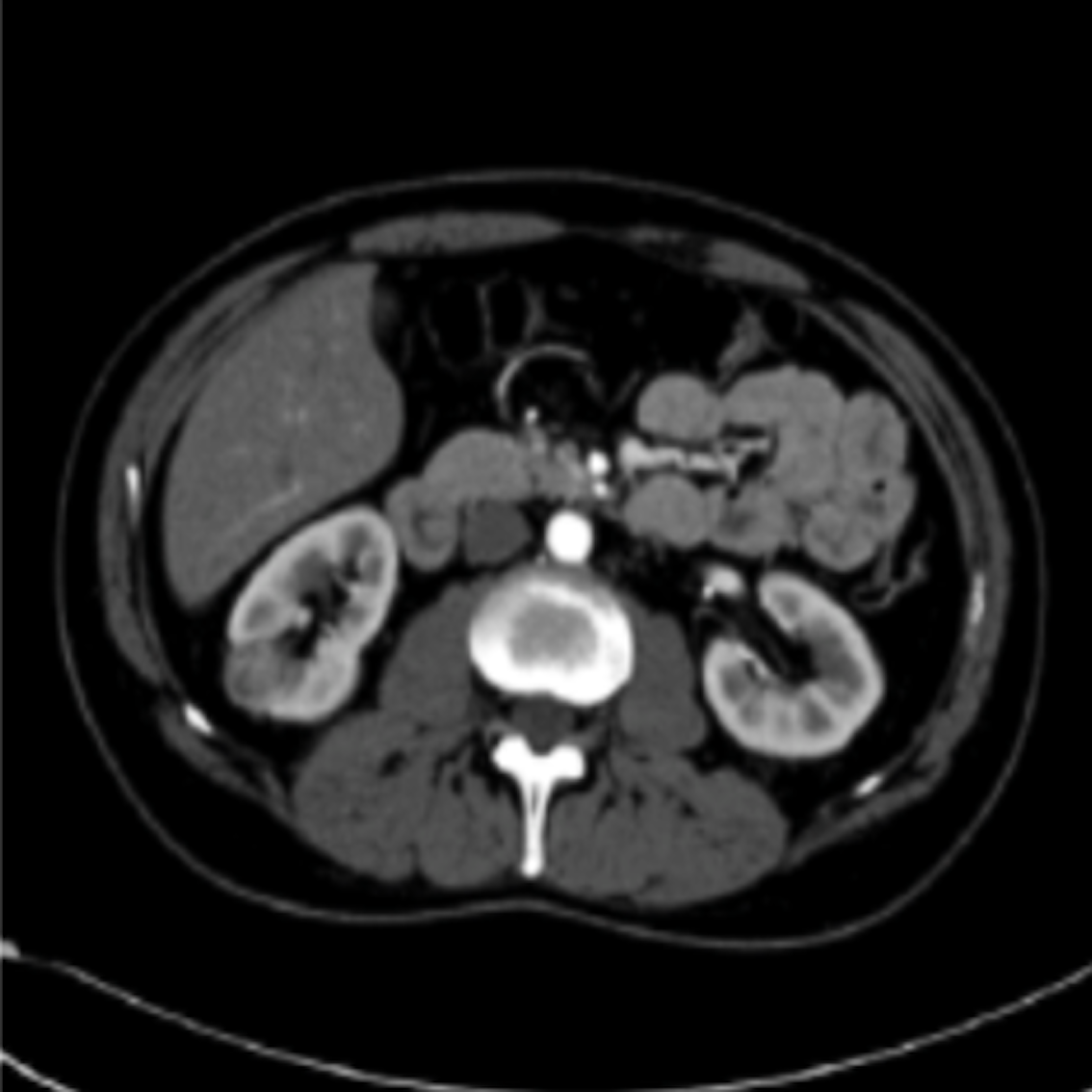}}
     \end{subfigure}
    \caption{(a) COVID (b) Normal (c) Pneumonia (d) Kidney CT scan.}
    \label{fig:covid_data}
\end{figure}

\begin{figure}[!ht]
    \centering
    \begin{subfigure}[]
    {\includegraphics[width=0.15\textwidth, height =0.12\textwidth]{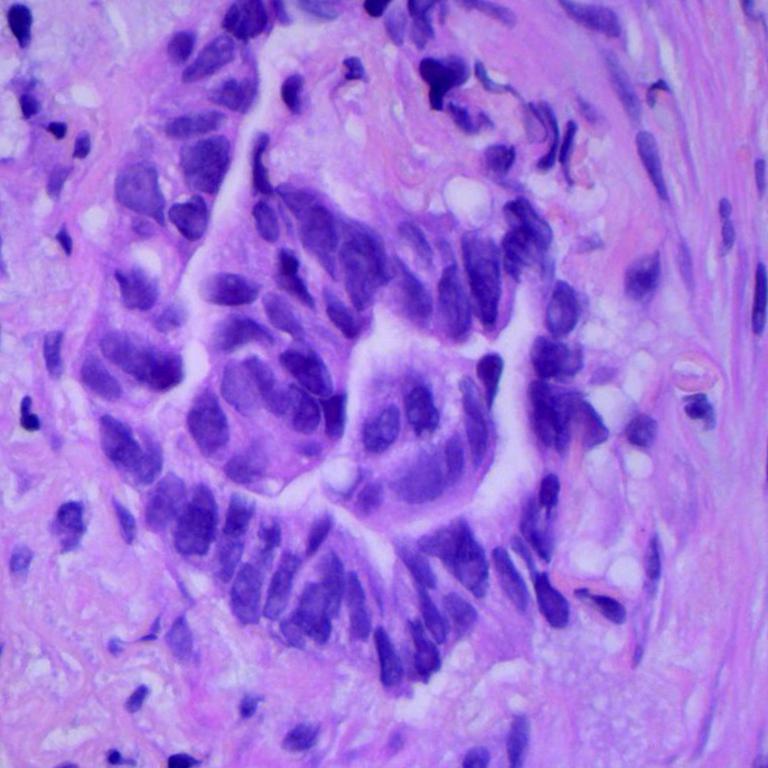}} 
    \end{subfigure}
    \begin{subfigure}[]
    {\includegraphics[width=0.15\textwidth, height =0.12\textwidth]{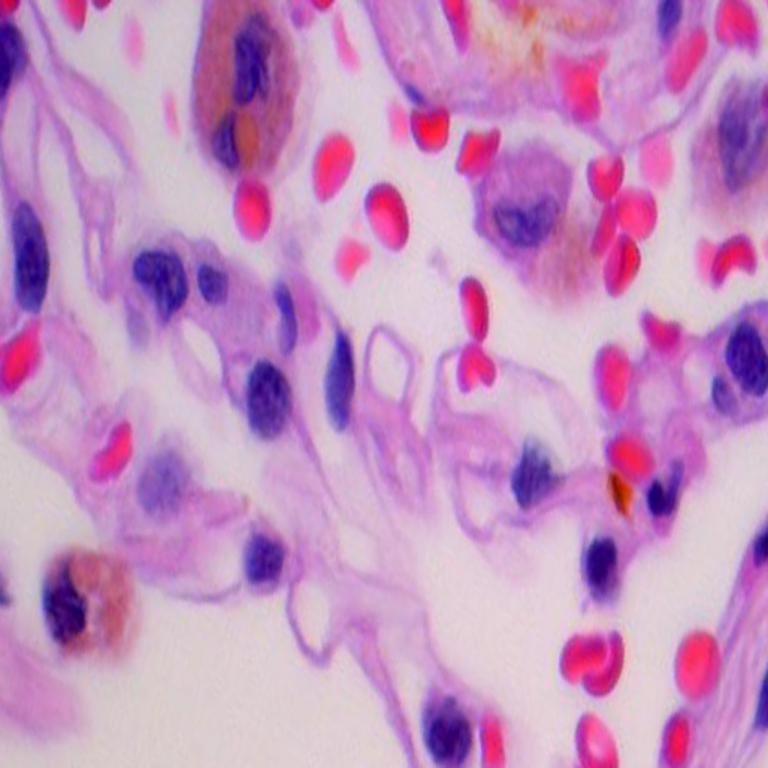}} 
    \end{subfigure}
    \begin{subfigure}[]
    {\includegraphics[width=0.15\textwidth, height =0.12\textwidth]{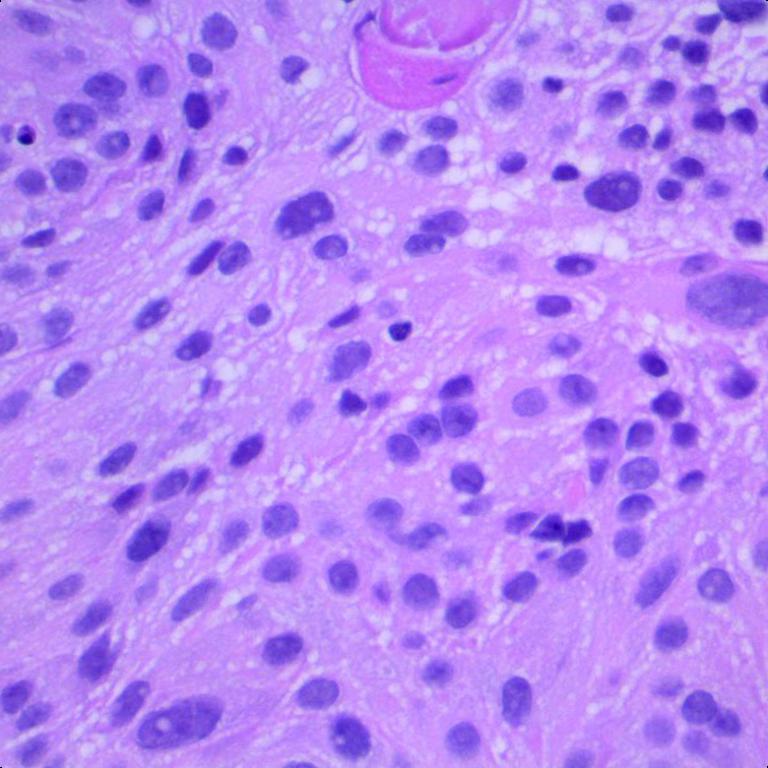}}
    \end{subfigure}
    \begin{subfigure}[]
    {\includegraphics[width=0.15\textwidth, height =0.12\textwidth]{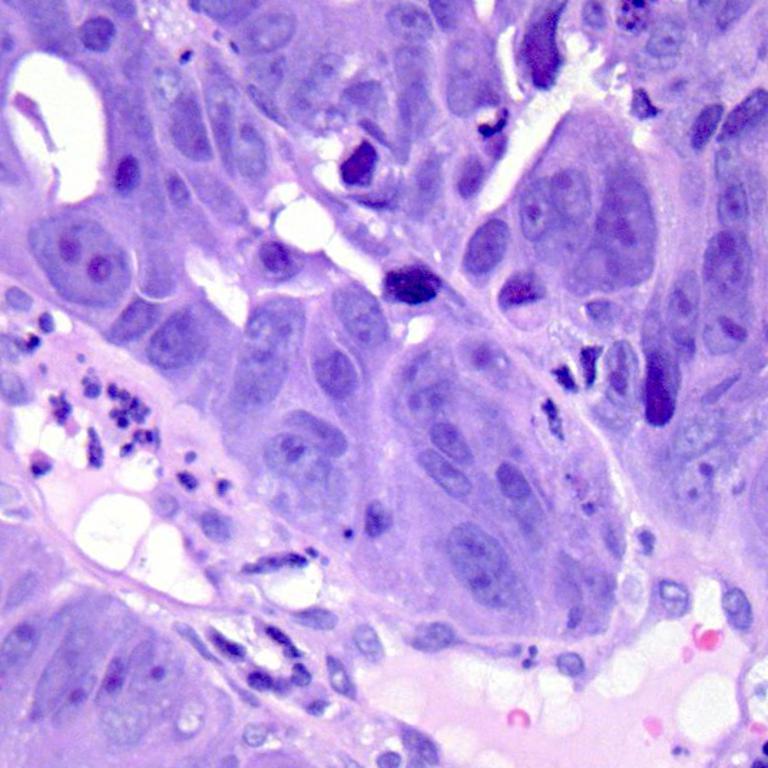}} 
    \end{subfigure}
    \begin{subfigure}[]
   {\includegraphics[width=0.15\textwidth, height =0.12\textwidth]{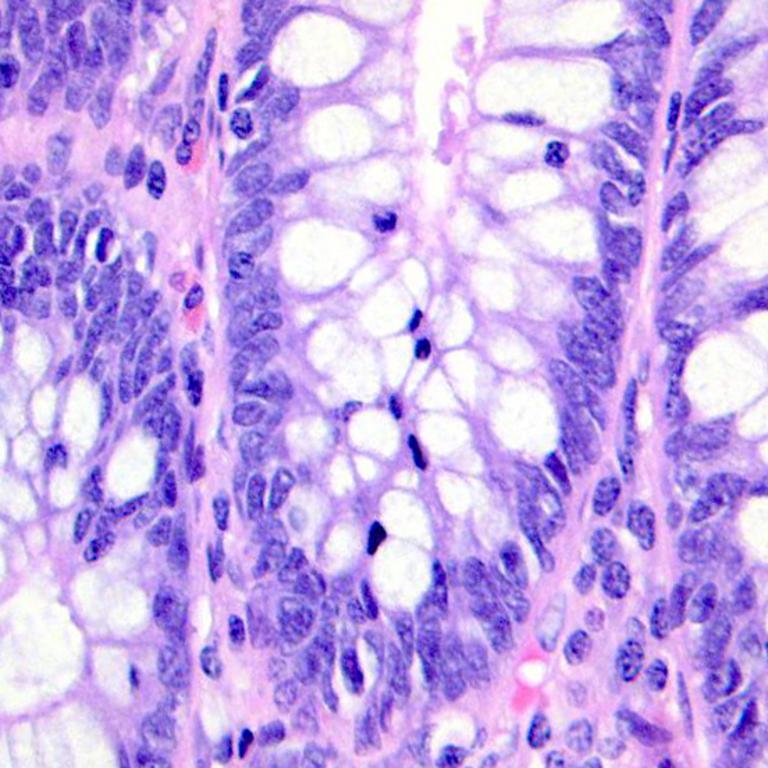}} 
   \end{subfigure}
    \caption{(a) Lung Adenocarcinomas (b) Lung squamous cell carcinomas (c) Lung benign  (d) Colon Adenocarcinomas (e) Colon Benign. }
    \label{fig:lc25000}
\end{figure}

In this paper, four publicly available medical image datasets have been used. For evaluation on grayscale images: COVID-19 chest X-ray~\footnote{https://www.kaggle.com/prashant268/chest-xray-covid19-pneumonia}, KiTS19 \cite{heller2020state} and for RGB images: LC25000 dataset \cite{borkowski2019lung} have been used. For a fair comparison, we used the same data splits as being used in \cite{ heller2020state, borkowski2019lung}. COVID-19 chest X-ray, consists of 6432 images and have three classes: normal, pnuemonia and COVID-19 as shown in Figure \ref{fig:covid_data}. We divided it into stratified splits of 80\% train, 10\% validation and 10\% test set. KiTS19 comprises of 300 gray scale abdominal scans of kidney patients with average of 216 slices (highest slice number is 1059) as shown in Figure \ref{fig:covid_data}(d). The ground truth for segmentation was created by experts with each pixel  labeled as one of three classes: background, kidney or tumor. For validation three-fold cross validation has been used on 120 scans with their slices respectively.
LC25000 dataset \cite{borkowski2019lung} comprises of 25000 histopathlogical images with 5 classes as shown in Figure. \ref{fig:lc25000}. There are 5000 images per class. We divided LC25000 with the ratio of 80:10:10 for train, validation and test set,respectively.


\subsection{Implementation Details}
SGD optimizer with a learning rate of $1e^{−4}$ was adopted and all models were trained for 100 epochs with batch size of 16. The cross entropy criterion is used. The codebase was setup in PyTorch framework and is available at github link \footnote{https://github.com/aleemsidra/Augmentation-Based-Generalized-Enhancement}.

\subsection{Statistical Evaluation}
\subsubsection{Evaluation on gray scale dataset}
The proposed approach effectiveness is evaluated by conducting different experiments. Firstly, to evaluate the strength of the proposed enhancement technique, we used gray scale datasets: COVID-19 chest X-ray and KiTS19 \cite{heller2020state}. The performance is then compared with the other existing techniques. For COVID-19 chest X-ray, classification is performed and the accuracy score is used as evaluation criterion. For KiTS19 \cite{heller2020state}, segmentation is performed and performance is accessed using  dice coefficient are used as an evaluation metric. In contrast to exisitng methods,  our method achieved highest accuracy and dice coefficient score with COVID-19 chest X-ray and KiTS19, respectively.



\begin{table}[!hb]
\caption{Accuracy comparison on the COVID-19 chest X-ray dataset}
\label{tab:tab1}
{\resizebox{\textwidth}{!}{\begin{tabular}{|l|c|c|c|c|c|c|c|}
\hline
Model & ResNet18  & ResNet50 & ResNet101 &  VGG16 & VGG19 & Inception &  DLH\_COVID \\ 
\hline
Existing & 0.9471  &  0.9346 & 0.9269 & 0.9486 & 0.9330 & 0.9144 & 0.9611 \\ 
\hline
\textbf{Proposed} & \textbf{0.9642}  & \textbf{0.9580} & \textbf{0.9486}  & \textbf{0.9611} & \textbf{0.9517} & \textbf{0.9315} & \textbf{0.9626} \\ 
\hline
\end{tabular}}}
\end{table}
With our enhancement approach, state-of-the-art DL architectures achieved superior performance to that achieved with un-enhanced data as evident from Table \ref{tab:tab1} and Table \ref{tab:tab2}. Thus, our enhancement approach aids the diagnostic performance of state-of-the-art DL architecture. 


\begin{table}[!ht]
\begin{center}
\caption{Dice coefficent for segmentation on  KiTS19 dataset}
\label{tab:tab2}
\begin{tabular}{|l|c|c|c|c|c|c|}
\hline
Model& U-Net & 3D FCN  & VB-Net  &  3D U-Net & MIScnn & \textbf{Proposed} \\ 
\hline
Kidney & 0.9663 &  0.9805 & 0.9740 & 0.9743 &  0.9994 & \textbf{0.9998} \\ 
\hline
Tumor & 0.7778 &  0.8370  & 0.7890  & 0.8558 &  0.9319 & \textbf{0.9411} \\ 
\hline

Background & - &  -  & -  & - &  0.6750 & \textbf{0.6820} \\ 
\hline

\end{tabular}
\end{center}
\end{table}
\subsubsection{Evaluation on RGB dataset}
To further validate the effectiveness of the proposed method, we extended our work to RGB datasets. For this purpose, we used the LC25000 dataset \cite{borkowski2019lung}. To best of our knowledge, \cite{mangal2020convolution} reported the highest accuracy with LC25000 dataset. So, we used its architecture to evaluate our proposed technique. With RGB datasets too, our method outperformed other existing techniques and achieved the highest accuracy as shown in Table \ref{tab:tab3}. It justifies our claim of proposing a generalized contrast enhancement method for different types of datasets i.e. gray scale and RGB.  


\begin{table}[!ht]
\begin{center}
\caption{Accuracy comparison of DL frameworks on the LC25000 dataset}
\label{tab:tab3}

\begin{tabular}{|l|c|c|c|c|c|c|}
\hline
Model& RF  & Resnet50  & CNN 
 &  CNN & CNN & \textbf{Proposed} \\ 
\hline
Lung & - &  - & 0.9720 & 0.9720 &  0.9789 & \textbf{0.9844} \\ 
\hline
Colon & 0.8530 &  0.9391  & -  & 0.9720 &  0.9616 & \textbf{0.9688} \\ 
\hline
\end{tabular}
\end{center}
\end{table}




\subsection{Visual Evaluation}
The visual analysis is equally important as clinicians will eventually use images for diagnosis. Firstly, to show the impact of the random hyperparameters, we selected the end points of brightness and contrast set range as discussed above in section \ref{intro}. The impact is shown below in Figure. \ref{fig:fig4}.

\begin{figure}[!ht]
\settoheight{\tempdima}{\includegraphics[width=.1\linewidth]{example-image-a}}%
\centering\begin{tabular}{c@{ }c@{ }c@{ }c@{ }c@{ }c@{ }}
 & \textbf{Original} & $\alpha$ = 1.15, b  = -0.1 &$\alpha$ = 1.15, b = 0.4 & $\alpha$ = 1.35,b = -0.1 & $\alpha$ = 1.35, b = 0.4\\
\rowname{}& 
\includegraphics[width=0.15\textwidth, height =0.12\textwidth]{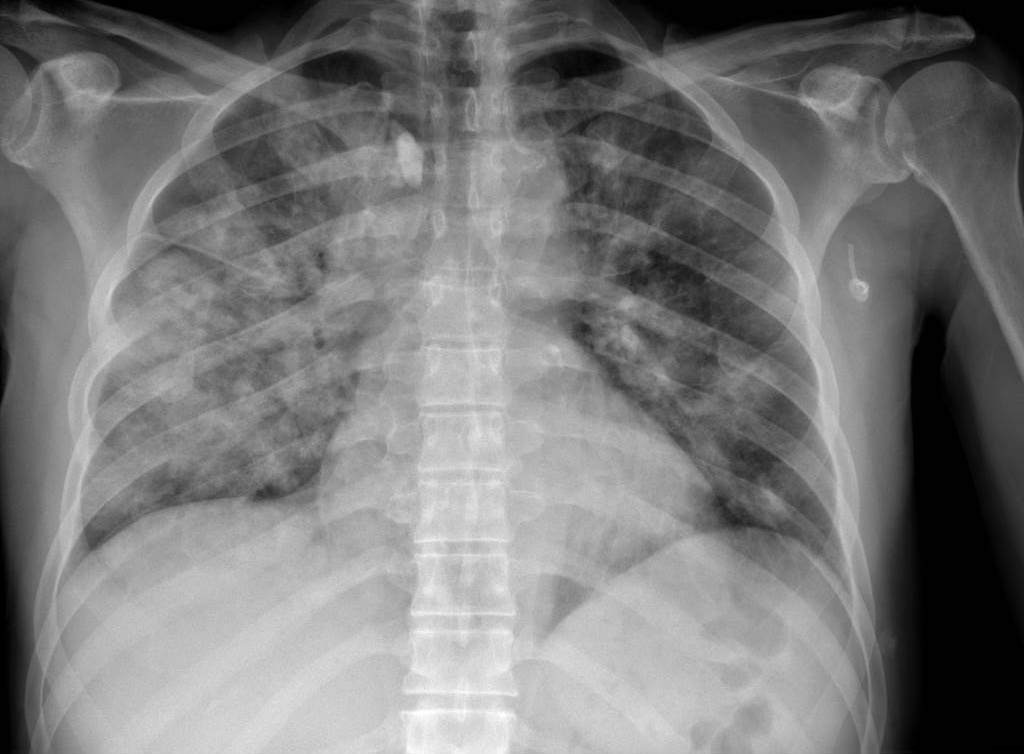}&
\includegraphics[width=0.15\textwidth, height =0.12\textwidth]{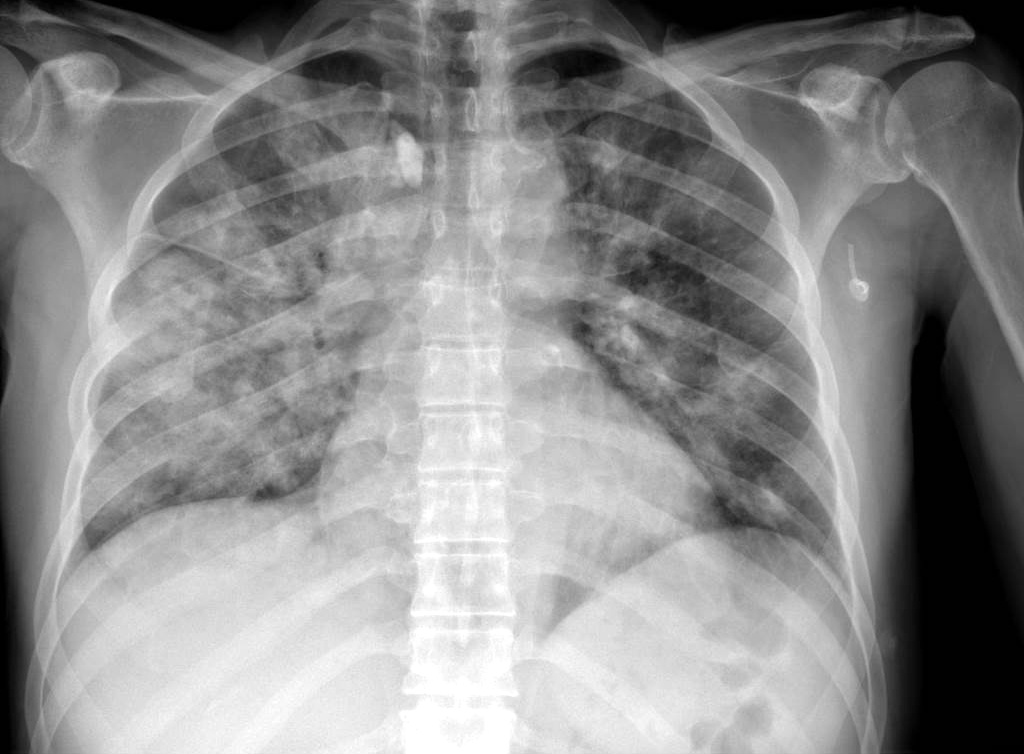}&
\includegraphics[width=0.15\textwidth, height =0.12\textwidth]{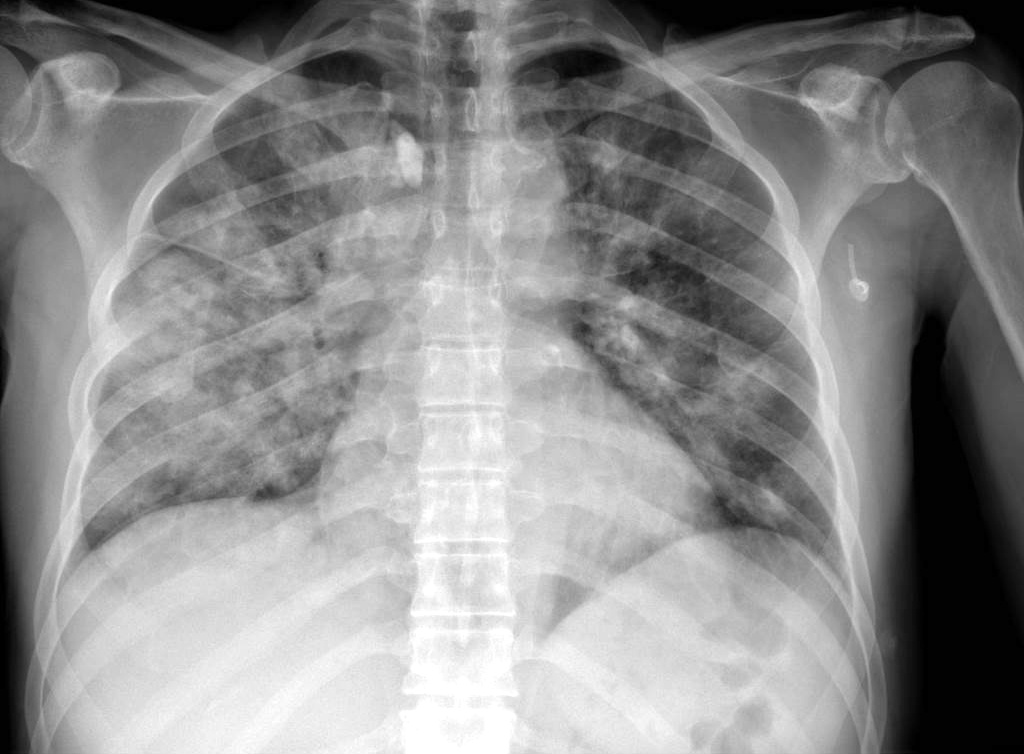}&
\includegraphics[width=0.15\textwidth, height =0.12\textwidth]{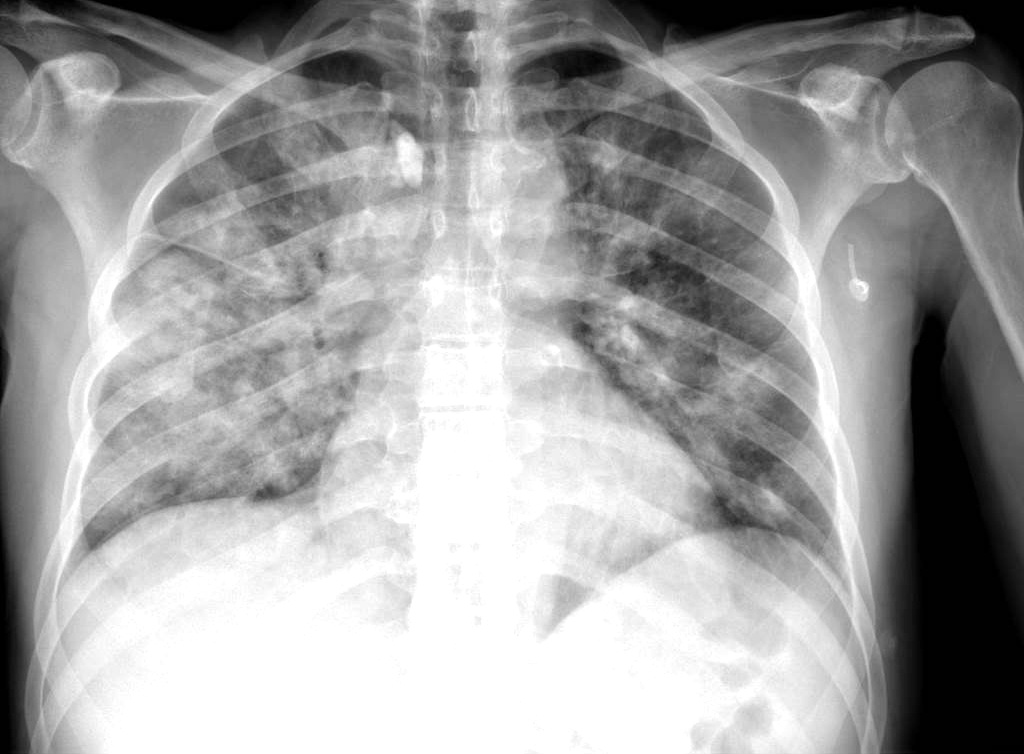}&
\includegraphics[width=0.15\textwidth, height =0.12\textwidth]{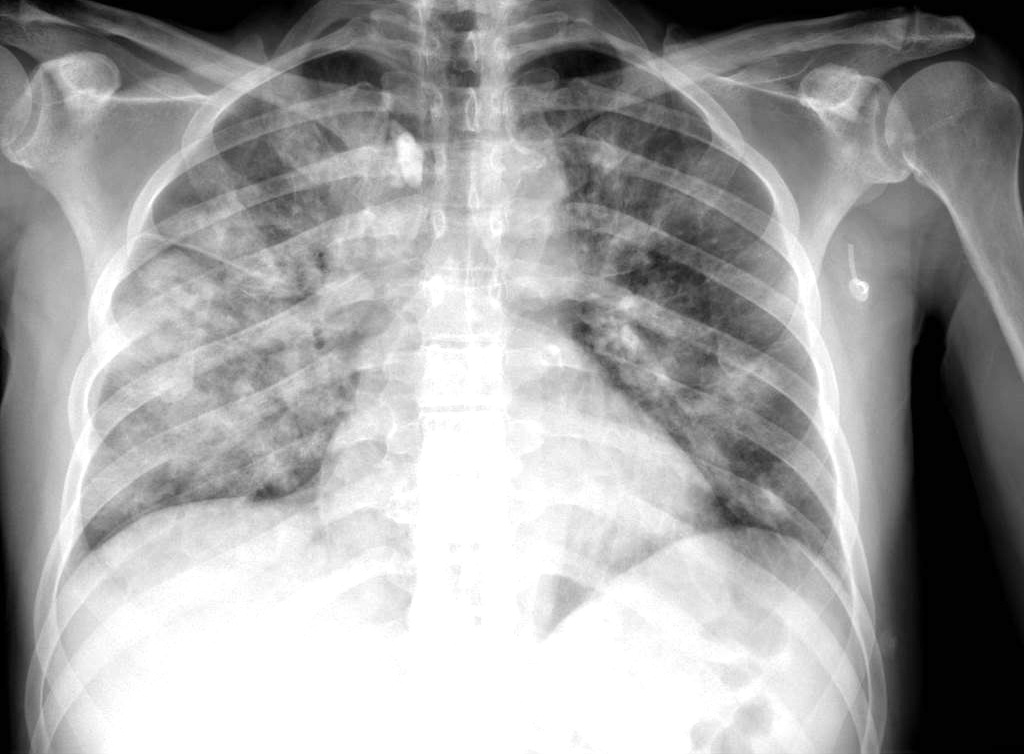}\\
\rowname{}&
\includegraphics[width=0.15\textwidth, height =0.12\textwidth]{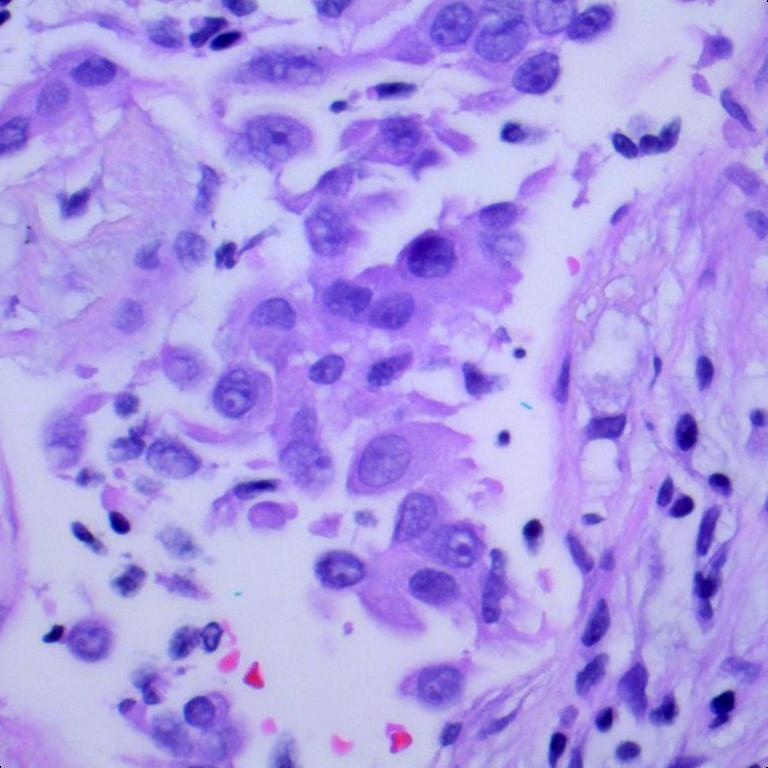}&
\includegraphics[width=0.15\textwidth, height =0.12\textwidth]{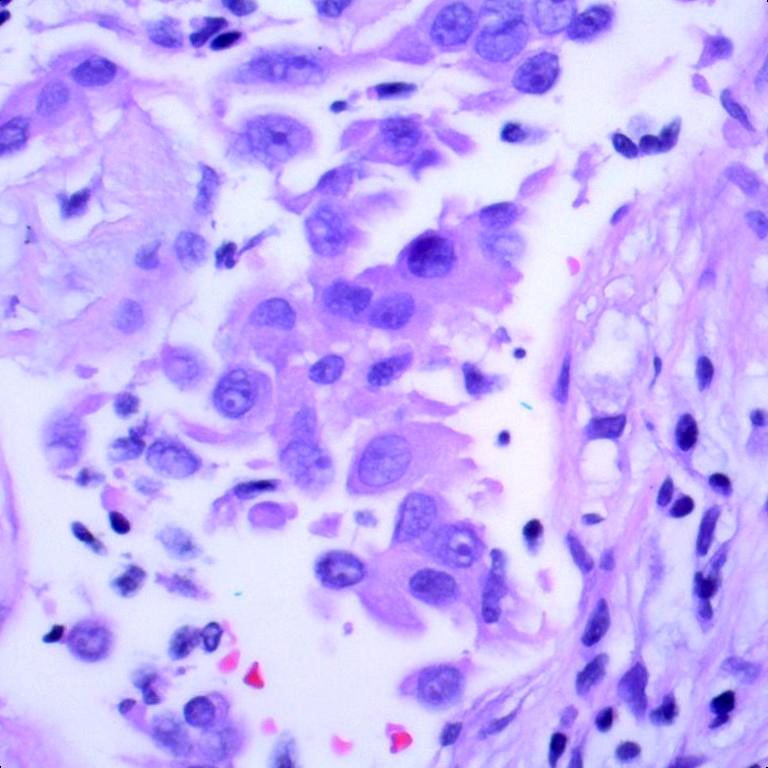}&
\includegraphics[width=0.15\textwidth, height =0.12\textwidth]{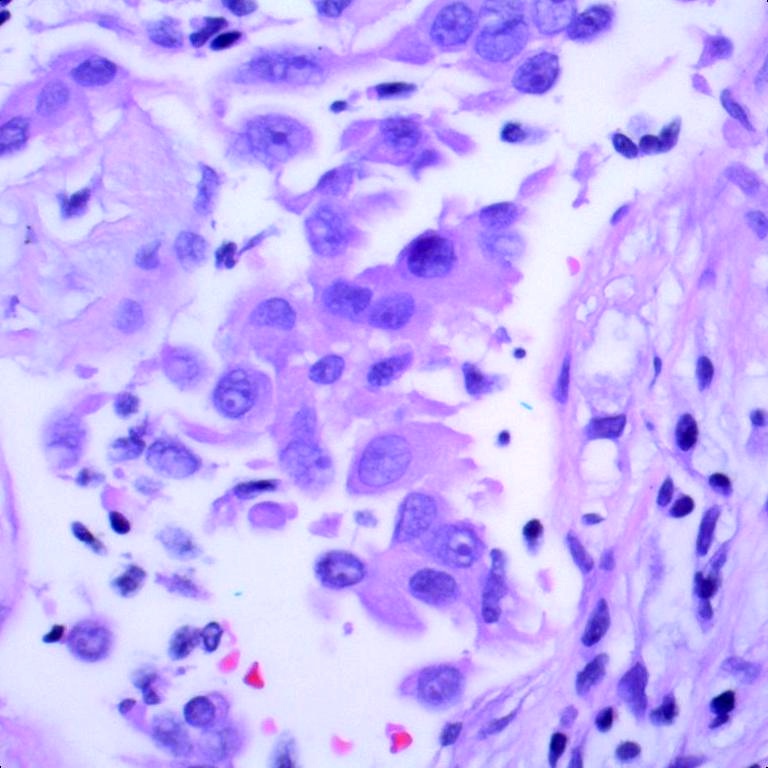}&
\includegraphics[width=0.15\textwidth, height =0.12\textwidth]{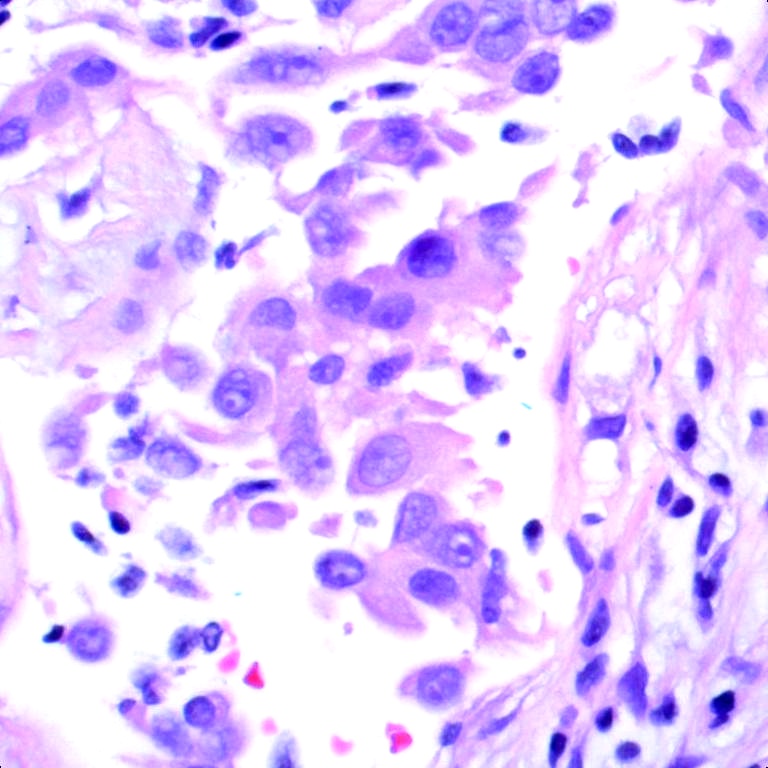}&
\includegraphics[width=0.15\textwidth, height =0.12\textwidth]{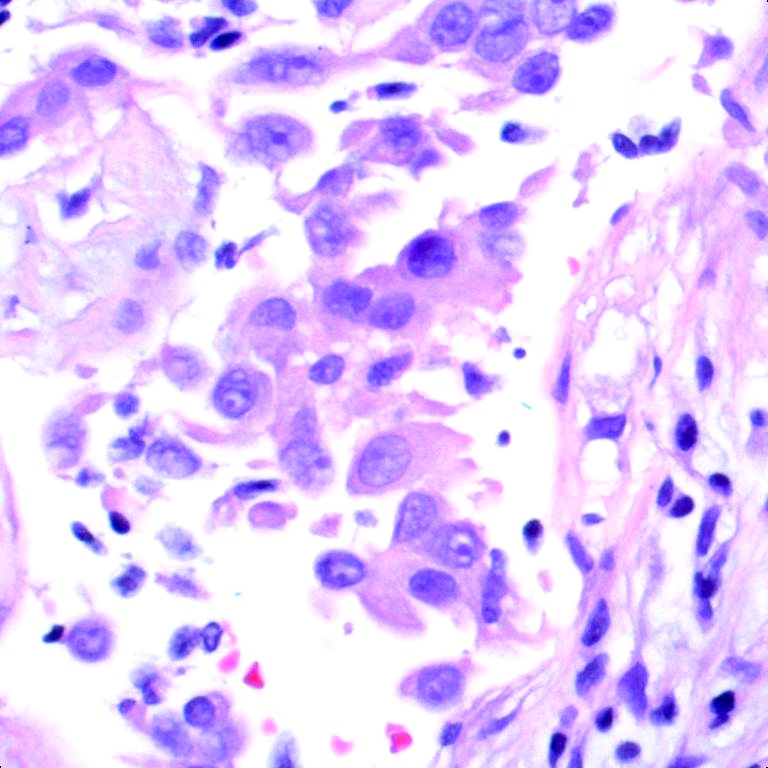}\\[-1ex]\\
\end{tabular}
\caption{Ablation study to evaluate the impact of different hyper-parameters on data quality. The first column corresponds to original image and rest of the columns show imapct of different $\alpha$ and $\beta$ values on image quality. }%
\label{fig:fig4}
\end{figure}

\begin{figure}[!ht]
\settoheight{\tempdima}{\includegraphics[width=.1\linewidth]{example-image-a}}%
\centering\begin{tabular}{@{}c@{ }c@{ }c@{ }c@{ }c@{ }c@{ }}
 & \textbf{COVID} & \textbf{Normal} &\textbf{Kidney} & \textbf{Squamous} & \textbf{Benign} \\
\rowname{Input}& 
\includegraphics[width=0.15\textwidth, height =0.12\textwidth]{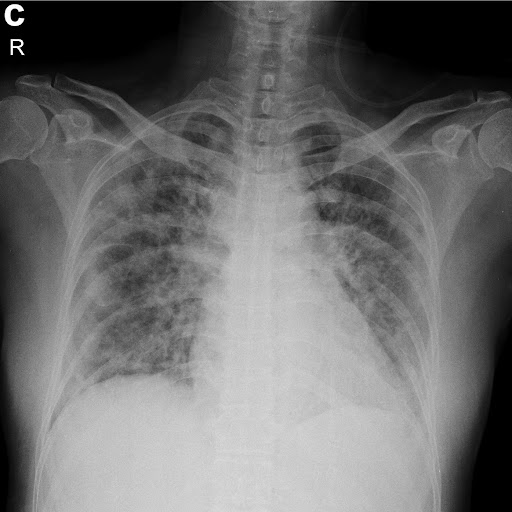}&
\includegraphics[width=0.15\textwidth, height =0.12\textwidth]{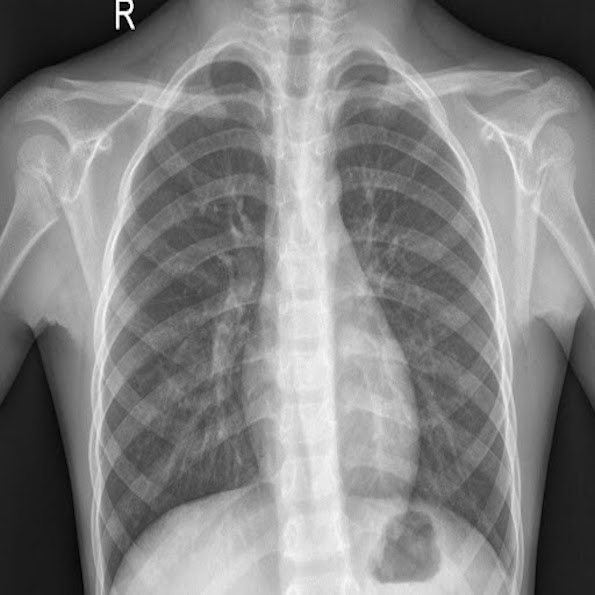}&
\includegraphics[width=0.15\textwidth, height =0.12\textwidth]{Images/kidney.png}&
\includegraphics[width=0.15\textwidth, height =0.12\textwidth]{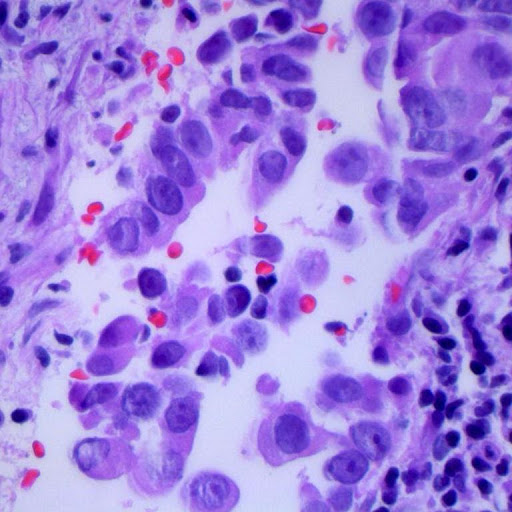}&
\includegraphics[width=0.15\textwidth, height =0.12\textwidth]{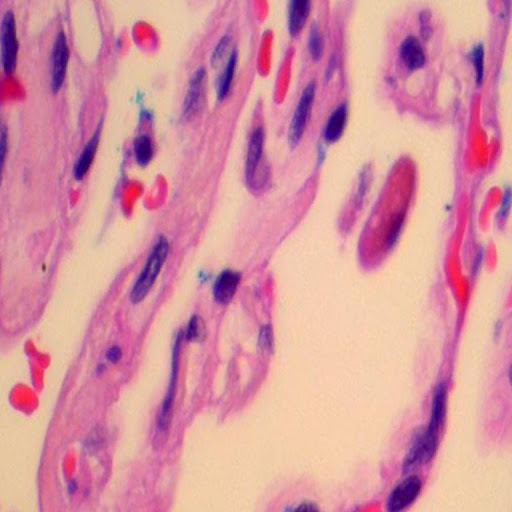}\\
\rowname{Enhanced}&
\includegraphics[width=0.15\textwidth, height =0.12\textwidth]{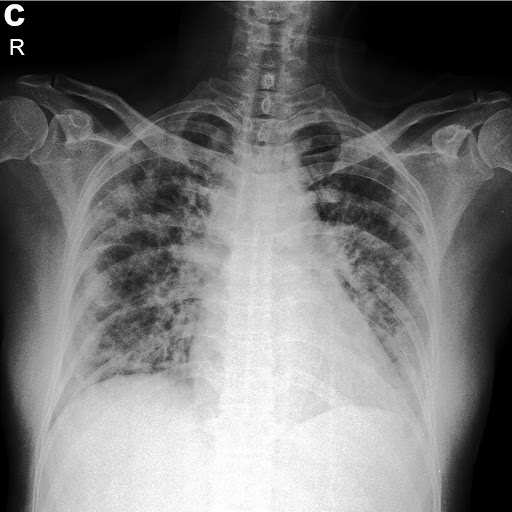}&
\includegraphics[width=0.15\textwidth, height =0.12\textwidth]{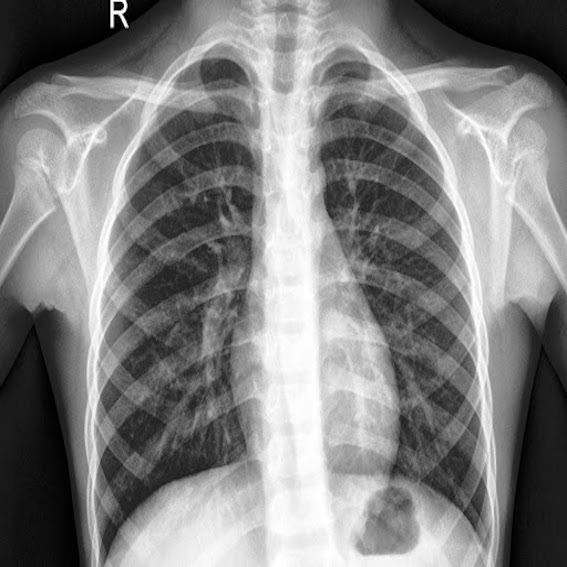}&
\includegraphics[width=0.15\textwidth, height =0.12\textwidth]{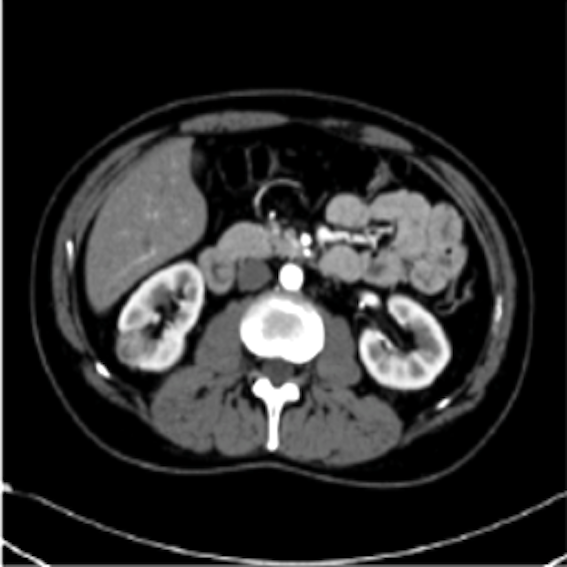}&
\includegraphics[width=0.15\textwidth, height =0.12\textwidth]{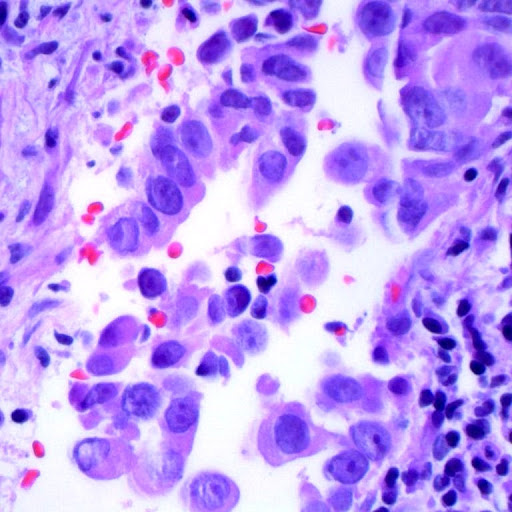}&
\includegraphics[width=0.15\textwidth, height =0.12\textwidth]{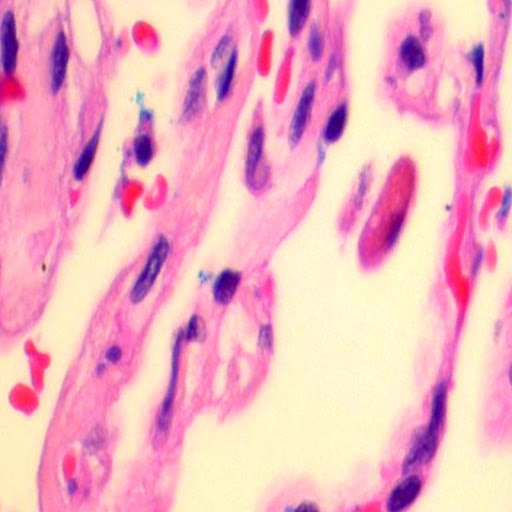}\\[-1ex]\\
\end{tabular}
\caption{Visual results of enhancement: row 1 corresponds to input images, row 2 corresponds to enhanced images. }%
\label{fig:fig6}
\end{figure}

Further, to show the effectiveness of our enhancement method, we compare the original images with the images enhanced by our method as shown in Figure \ref{fig:fig6}. It is evident that quality is significantly improved by our method. 

\section{Conclusion}
In this paper, we proposed a new enhancement technique based on data augmentation that improves medical image quality from brightness and contrast perspective. It uses a random hyperparameters selection from the evaluated set of values. The proposed approach addressed the limitations of existing approaches including DL complex architecture and fixed hyperparameters based enhancement. Furthermore, to evaluate the generalization capability of the proposed approach, we used four datasets including grayscale and RGB with a wide variety of networks for both classification and segmentation tasks. The proposed approach, despite being a simple and easy to implement, outperforms existing approaches using state-of-the-art networks. To reproduce the results we made our code publicly available \url{https://github.com/aleemsidra/Augmentation-Based-Generalized-Enhancement}.

\section*{Acknowledgement}
This research was supported by Science Foundation Ireland under grant numbers 18/CRT/6183 (ML-LABS Centre for Research Training),18/CRT/6223, SFI/12/RC/2289{\_}P2 (Insight SFI Research Centre for Data Analytics),  13/RC/$2094\_P2$ (Lero SFI Centre for Software ) and 13/RC/$2106\_P2$ (ADAPT SFI Research Centre for  AI-Driven Digital Content Technology). For the purpose of Open Access, the author has applied a CC BY public copyright licence to any Author Accepted Manuscript version arising from this submission.

\bibliographystyle{apalike}

\bibliography{imvip}

\end{document}